\documentclass{kluwer} 
\newdisplay{guess}{Conjecture}
\def\note #1]{{\bf #1]}} 
\def\be{\begin{equation}}
\def\ee{\end{equation}} 
\def\bearr{\begin{eqnarray}} 
\def\eearr{\end{eqnarray}}
\def\barr{\begin{array}}

\def\aap{\it Astron. Astrophys. \rm}

\def\apj{\it Astrophys. J. \rm}
\begin{document}
\begin{article}
\begin{opening}

\title{EMPIRICAL ESTIMATE OF p-MODE FREQUENCY SHIFT FOR SOLAR CYCLE 23}
\author{Kiran \surname{Jain}\email{kiran@uso.ernet.in},
S. C. \surname{Tripathy}\email{sushant@uso.ernet.in}, 
A. \surname{Bhatnagar}\email{arvind@uso.ernet.in} and Brajesh \surname{Kumar}\email{brajesh@uso.ernet.in}}
\institute{Udaipur Solar Observatory, A unit of Physical Research Laboratory, Off Bari Road,
 Dewali, P.  B.  No.  198, Udaipur 
313001, India} 
\runningtitle{EMPIRICAL ESTIMATE OF FREQUENCY SHIFT}
\runningauthor{Jain, Tripathy, Bhatnagar and Kumar }
\begin{ao}
Dr. Kiran Jain \\ Udaipur Solar Observatory\\ Off Bari Road,
Dewali, P.  B.  No.  198, \\ Udaipur - 313001, \\ India \\
e-mail:kiran@uso.ernet.in
\end{ao}
\begin{abstract}

We have obtained empirical relations between the p-mode
frequency shift and the {\it change} in solar activity indices.
The empirical relations are determined on the basis of
frequencies obtained from BBSO and GONG stations during solar cycle
22.  These relations are applied to estimate the change in mean
frequency for the cycle 21 and 23. A remarkable agreement
between the calculated and observed  frequency shifts for the
ascending phase of cycle 23, indicates that the derived
relations are independent of epoch and do not change
significantly from cycle to cycle.  We propose that these
relations could be used to estimate the shift in p-mode
frequencies for past, present and future solar activity cycles,
if the solar activity index is known. The maximum frequency
shift for cycle 23 is estimated to be 265 $\pm$ 90~nHz,
corresponding to the predicted maximum smoothed sunspot number
118.1 $\pm$ 35.

\end{abstract}
\date{\today}
\end{opening}

\section{Introduction}

It is now well established that the solar p-mode oscillation
frequencies vary with solar activity. The first evidence of
this effect came from the analysis of low degree acoustic
frequencies derived from solar irradiance data by the Active
Cavity Radiometer instrument on board {\it Solar Maximum
Mission} satellite \cite{wn85}. Using the Doppler velocity data,
\inlinecite{palle89} also showed that the frequency shifts are
well correlated with solar activity cycle and obtained a
shift of 0.44 $\pm$ 0.06~$\mu$Hz between the minimum and maximum of
solar cycle 21.  Later, several other authors have extended
these studies to different epochs with new and improved data
sets for intermediate (\opencite{wod91}; \opencite{bb93}) and
low degree modes (Jim\'{e}nez-Reyes {\it et al.}, 1998 and {\it
references therein}).  
 A consistent and continuous data set
of intermediate degree p-mode frequencies is recently made
available from the Global Oscillation Network Group (GONG) for
the period  May 1995 to October 1998, with an interval of 108
days.  Using some of these data sets between August 1995 to
August 1997, \inlinecite{ab99} obtained a decrease of
0.06~$\mu$Hz in the mean frequency during the descending phase
of the solar cycle 22 and an increase of 0.04~$\mu$Hz during the
ascending phase of the cycle 23.  \inlinecite{howe99}, using a
subset of GONG frequencies, also demonstrated that the p-mode
frequencies vary with solar activity cycle.

	 The prime motivation of this work is to derive 
empirical relations between the shift in frequency and {\it
change} in the level of the activity indices. The derived relations
could be used to estimate the frequency shift for past and
future solar cycles. In this paper, we have used the
{\it change} in activity indices corresponding to the same
epoch as the frequency shifts instead of the {\it actual}
value of activity. We find that the correlation between change
in activity and frequency shifts is better as compared when
the actual value of the activity indices were used. Using a similar approach
\inlinecite{rhodes93} showed that the frequency shifts between
1981 and 1989 are correlated with the {\it change} in
various activity indices; e.g., the sunspot number, sunspot area,
and irradiance measurements.

\section{Observational data}
The mode frequencies for cycle 22 in the intermediate degree
range are available from the Big Bear Solar observatory (BBSO),
LOWL instrument, South Pole expeditions and GONG project. As
shown by \inlinecite{jain99}, frequencies derived from South
Pole observations are systematically higher as compared with
other three data sets.  Similarly frequencies derived from LOWL
instrument are from one year power spectra, which may not be
valid for the study of short period solar cycle variation.
Thus, in this study we use p-mode frequencies obtained from BBSO
and GONG, for the period 1986 to 1996; these are summarised in
Table~I. It may be noted that we have used only the
continuous and independent data sets from the GONG network.
As it was  shown by \inlinecite{lw90} that the mode frequency shift strongly depends on
the frequencies, we have used only the common modes available 
between GONG and BBSO data sets. This selection criteria generated a total number
of 412 common modes in the frequency range of 1500 and 3500
$\mu$Hz and $\ell$ between 5 to 99. Further, we have
categorised the data sets into two groups according to the data
source as shown in Table~I.

\begin{table*} \caption[]{Frequency Data Sets}
\label{tab1}
\begin{tabular}[]{llcc}
\hline 
\noalign{\smallskip} 
Data Set & \multicolumn{1}{c}{Epoch} & Extent & Number of Modes \\
& & (days) & (in 0 $\leq$ $\ell$ $\leq$ 100) \\
\noalign{\smallskip} 
\hline
BBSO86 & Mar-Aug 1986 & 131 &  1095\\ 
BBSO88 & Mar-Sep 1988 & 183 &  1095\\ 
BBSO89 & Mar-Sep 1989 & 182 &  1095\\  
BBSO90 & Mar-Sep 1990 & 180 &  1095\\
GM2 & 7 May-22 Aug 1995 & 108 &  1079\\
GM5 & 23 Aug-8 Dec 1995 & 108 &  1078\\
GM8 & 9 Dec-25 Mar 1996 & 108 & 1579 \\
GM11 & 26 Mar-11 July 1996 & 108 &  1143\\
GM14 & 12 July-27 Oct 1996 & 108 & 1055 \\
\noalign{\smallskip}
\hline
\end{tabular} 
\end{table*}

\section{Analysis and results}
The mean frequency shift is calculated by taking simple
difference between any two data sets chosen from the same group.
With 9 frequency data sets (see Table I), the possible
combinations (choosing any two at a time, out of the total data
sets from an individual group) produced 16 values of the mean
frequency shifts. In order to investigate how the frequency
shifts are correlated with the {\it change} in the level of activity
indices and to derive empirical relations between them, we have
used seven different activity indices representing the magnetic
and radiative indices. These indices are: R$_I$,  unsmoothed
International sunspot number obtained from the Solar Geophysical
Data (SGD); KPMI, Kitt Peak Magnetic Index from Kitt peak full
disk magnetograms; MPSI, Magnetic Plage Strength Index from
Mount Wilson magnetograms \cite{ulr91}; FI, total flare index
from SGD and Ata\c{c} (1999); He~I,
equivalent width of Helium~I 10830\AA \ line \cite{har84},
averaged over the whole disk from Kitt peak;   F$_{10}$,
integrated radio flux at 10.7 cm from SGD, and R$_s$,  smoothed
International sunspot number obtained from SGD.  A mean value
for each activity index was computed, for the epoch
corresponding to the actual frequency interval.

To study the relative variation in the mean frequency shift 
$\delta\nu$ with the change in activity index $\delta i$, we assume a 
linear relationship of the form:
\begin{equation}
\delta\nu  =  a~\delta i + b ,
\end{equation}
where slope $a$ and  
intercept $b$ are obtained by performing a linear least square fit and is shown in Figure~1. The solid line represents the '
best regression fit and confirms that the data sets are consistent with the assumption 
of a linear relationship. The bars represent 1$\sigma$ error in fitting.

\begin{figure}
\begin{center}
\leavevmode
\input epsf
\epsfxsize=3.0in \epsfbox{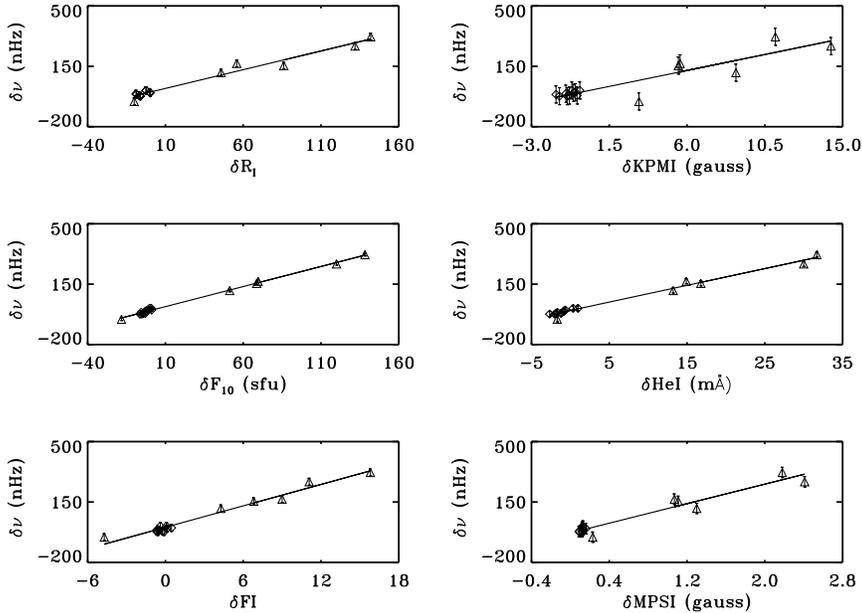}
\caption{Linear regression of mean frequency shift with change in 
activity indices. 
The solid line is the least square fit. The triangles represent BBSO data while squares are for GONG data.} 
\end{center}
\end{figure}

	The linear relationship given in Equation~(1) is further
tested by calculating the $\chi^2$, parametric Pearson's coefficient,
$r_P$, the rank correlation coefficient, $r_S$, and their
probabilities $P_p$ and $P_s$ respectively.  Table~II summarises
the correlation statistics for all the data sets included in the
fitting and shows that a positive correlation exists for all the
activity indices. We further note that the best correlation is
obtained for F$_{10}$, which confirms earlier results
(\opencite{bb93}; \opencite{ab99}) that the radiative indices
are better correlated with the frequency shifts. From the slope
$a$ and the gradient $b$ obtained from the regression fitting,
the following empirical relations between the shift, $\delta\nu$
and {\it change} in activity indices, $\delta i$ are formulated:

\begin{table*} \caption[]{Correlation statistics for Solar Cycle 22}
\label{tab2}
\begin{tabular}[]{lccccc}
\hline 
\noalign{\smallskip} 
Activity Index & $\chi^2$&$r_{P}$ & $P_{p}$ & $r_{S}$ & $P_{s}$ \\
\noalign{\smallskip} 
\hline
R$_I$ &13& 0.99 & 1.4E$-12$ & 0.97 & 4.7E$-$10 \\
KPMI &72& 0.91 & 7.1E$-07$ & 0.73 & 1.2E$-$03 \\
F$_{10}$ & 30&1.00 & 3.9E$-19$ & 0.99 & 8.1E$-$13 \\
He I  & 12&0.99 & 3.2E$-14$ & 0.93 & 1.9E$-$07 \\
FI & 52&0.98 & 5.0E$-12$ & 0.94 & 8.5E$-$08 \\
MPSI & 25&0.97 & 3.7E$-10$ & 0.81 & 1.4E$-$04 \\
R$_s$ &38& 0.99 & 1.0E$-13$ & 0.95 & 1.4E$-$08 \\
\noalign{\smallskip}
\hline
\end{tabular} 
\end{table*}
\begin{eqnarray}
\delta\nu & = &(2.44 \pm 0.18)~\delta R_I - (6.00 \pm 1.36) \\
\delta\nu  &= &(18.70 \pm 1.62)~\delta KPMI - (0.82 \pm 1.77) \\
\delta\nu  &= &(2.66 \pm 0.20)~\delta F_{10} - (4.67 \pm 1.44) \\
\delta\nu & =& (9.82 \pm 0.71)~\delta HeI + (0.83 \pm 1.68) \\
\delta\nu & = &(23.03 \pm 1.86)~\delta FI - (11.10 \pm 1.22) \\
\delta\nu & = &(159.01 \pm 11.90)~\delta MPSI - (36.10 \pm 1.86) \\
\delta\nu & =& (2.41 \pm 0.19)~\delta R_s - (0.48 \pm 1.68) 
\end{eqnarray}
where $\delta\nu$ is given in nHz and the change in
activity indices have their standard units. The second term in
the parenthesis indicates 1$\sigma$ error. As mentioned previously, 
these relations are valid in the frequency 
range of 1500 to 3500 $\mu$Hz and $\ell$ between 5 to 99.

\begin{figure}
\begin{center}
\leavevmode
\input epsf
\epsfxsize=4.5in \epsfbox{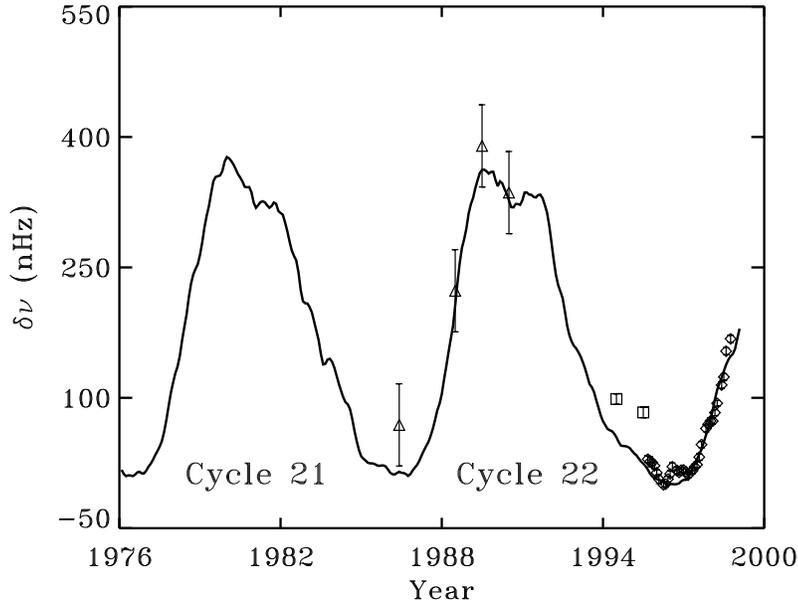}
\caption{Comparison of estimated and observed mean frequency
shift for cycle 21 through 23. The solid line is the mean shift
as calculated from Equation~(8).  The observed shifts are
represented by symbols; triangles for BBSO, squares for LOWL
instrument and diamonds for GONG. The vertical line within the
symbols represents the magnitude of 1$\sigma$ error in the
frequncy shift.}
\end{center}
\end{figure}

These relations are used to estimate the mean frequency shifts
for solar cycle 21 through 23 by taking the frequency data set GM2 as the
reference point.  As an example, the estimated frequency shift
obtained from Equation (8) for the smoothed sunspot number is
plotted in Figure~2 and yields a  shift of 0.37 $\pm$ 0.03
$\mu$Hz between the minimum and maximum of the solar cycle 21.
This value is consistent with the earlier results by
\inlinecite{palle89} and \inlinecite{els90}, who had obtained a
shift of 0.44 $\pm$ 0.06 $\mu$Hz and 0.46 $\pm$ 0.06 $\mu$Hz
respectively for the same period, using the Doppler velocity
data of low degree ($\ell$~ $\leq$~ 3) modes. From Figure~2, we
notice that a very good agreement exists between the calculated
values of $\delta\nu$ and the observed GONG frequencies
(diamonds), however a small difference for BBSO (triangles) and
LOWL (squares) is noticed. This  difference may be interpreted as
due to the use of different spectral lines in BBSO and LOWL
instruments \cite{howe99} or different data reduction techniques.

\subsection{Comparison between cycle 22 and 23}  

\begin{figure}
\begin{center}
\leavevmode
\input epsf
\epsfxsize=4.5in \epsfbox{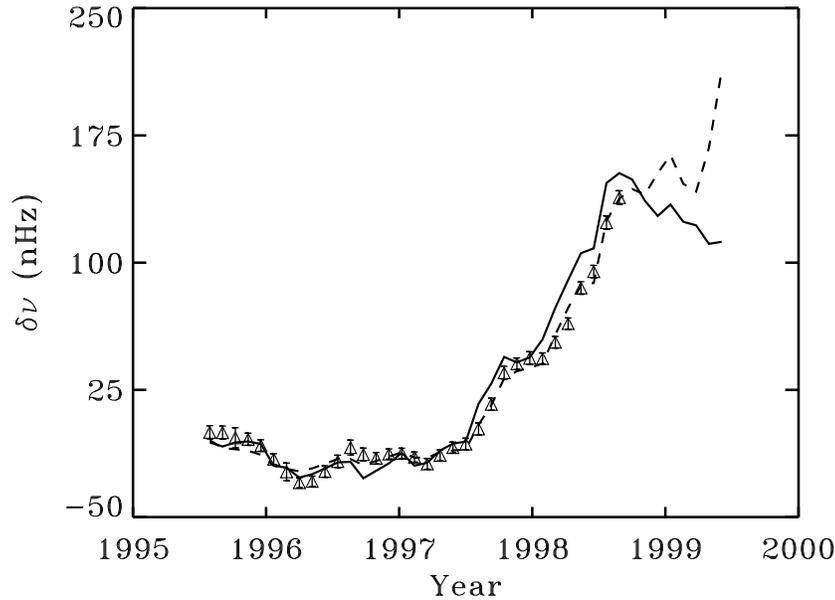}
\caption{The estimated and observed frequency shifts for 1995--1998.
The dashed line shows the shift,  as estimated
from the sunspot number (Equation~(2)) and the solid line for the 10.7 cm 
(Equation~(4)). 
The diamonds represent the observed frequency shifts from GONG data in the 
frequency range of 1500 and 3500
$\mu$Hz and $\ell$ between 5 to 99, 
where the mean shift is computed by taking GM2 as the reference.} 
\end{center}
\end{figure} 

To investigate the validity of empirical relations for different
cycles, we calculated the fitting parameters $a$ and $b$ for
cycle 23, using the available GONG data in the frequency range of 
1500-3500 $\mu$Hz and $\ell$ between 5 to 99.  These parameters for
solar cycle 22 and 23 are given in Table~III and a reasonable
agreeement is found for both the cycles.
 
\begin{table*} \caption[]{Values of gradient $a$ (in nHz/activity)
 and intercept $b$ (in nHz) obtained from least square fitting for solar 
cycle 22 and 23}
\label{tab3}
\begin{tabular}[]{lrrrr}
\hline 
\noalign{\smallskip} 
Activity Index&\multicolumn{2}{c}{cycle 22}&\multicolumn{2}{c}{cycle 23}\\
\rlcline{2-3}\rcline{4-5}
 & \multicolumn{1}{c}{$a$}  & \multicolumn{1}{c}{$b$} & 
 \multicolumn{1}{c}{$a$} & \multicolumn{1}{c}{$b$} \\
\noalign{\smallskip} 
\hline
R$_I$ & 2.44~$\pm$~0.18 & $-$6.00~$\pm$~1.36 & 2.01~$\pm$~0.07 & $-$4.20~$\pm$~0.56 \\
KPMI   & 18.70~$\pm$~1.62 & $-$0.82~$\pm$~1.77 & 22.6~$\pm$~0.79 & $-$1.86~$\pm$~0.56 \\
F$_{10}$ & 2.66~$\pm$~0.20 & $-$4.67~$\pm$~1.44 & 2.72~$\pm$~0.09 & $-$3.25~$\pm$~0.56 \\
HeI & 9.82~$\pm$~0.71 & $+$0.83~$\pm$~1.68 & 6.84~$\pm$~0.25 & $+$5.56~$\pm$~1.31 \\
FI  & 23.03~$\pm$~1.86 & $-$11.10~$\pm$~1.22 & 37.70~$\pm$~1.32 & $-$1.71~$\pm$~1.49 \\
MPSI & 159.01~$\pm$~11.90 & $-$36.10~$\pm$~1.86 & 107.00~$\pm$~4.17 & $-$15.12~$\pm$~2.06 \\

\noalign{\smallskip}
\noalign{\smallskip}
\hline
\end{tabular} 
\end{table*} 
  
 Therefore, we propose
 that the derived linear relations are independent of solar cycle and can be used to estimate
 the frequency shifts for past, present and future solar cycles. This is 
 illustrated in Figure~3, wherein we have plotted
 the calculated frequency shifts using Equations (2) and (4) and the observed  frequency shifts for cycle 23. The dashed and solid lines represent 
 the estimated $\delta\nu$, obtained from the relation for International 
sunspot number and 10.7 cm radio flux respectively, whereas the frequency shifts from the GONG data, with 
reference to GM2, are shown as triangles.  It is clear that the observed frequency shifts for cycle
23 are in close agreement with those obtained from the derived relations for cycle 22. 

\begin{figure}
\begin{center}
\leavevmode
\input epsf
\epsfxsize=4.5in \epsfbox{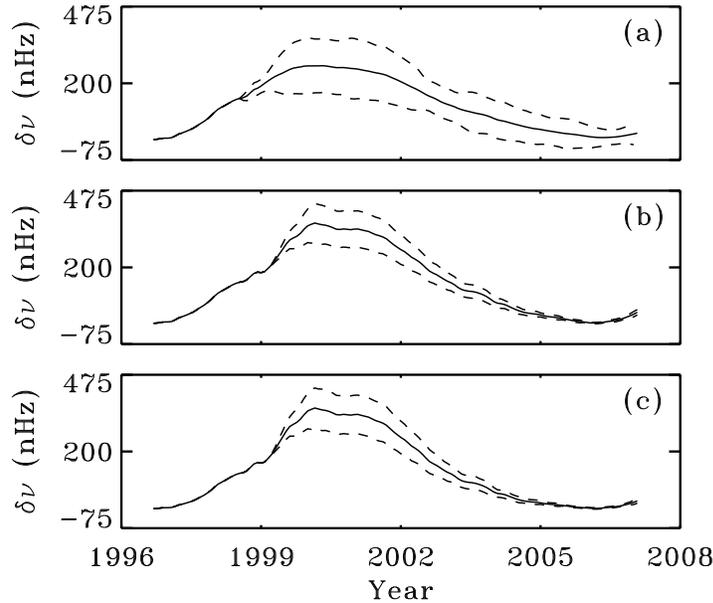}
\caption{(a) Estimated $\delta\nu$ for cycle 23 from (a) the predicted smoothed sunspot number from SGD 
web site (solid line) (b) predicted 
smoothed sunspot number and (c) 10.7 cm radio flux from MSFC web site.  The dashed lines 
 show  error 
in $\delta\nu$ due to the errors in predicted activity indices. } 
\end{center}
\end{figure} 

 In Figure~4, we have plotted the estimated frequency shifts for the current solar cycle 23, using 
predicted smoothed sunspot numbers and 10.7 cm flux. The solid line in Figure~4$a$
shows the predicted shift using R$_s$, as listed in Solar Geophysical 
Data web page (http://www.ngdc.noaa/gov/stp/stp.html). The dashed lines are 
the predicted  errors due to  error in R$_s$.  Each smoothed sunspot 
number represents the 
average of two adjacent 12-month running
mean of monthly means. The predicted value of sunspot number is based on the actual value of smoothed
sunspot number for February 1999, which utilises the values of the averaged monthly
means from August 1998 through August 1999. The predicted frequency shift for the solar cycle 23 is estimated to 
be 265 $\pm$ 90 nHz corresponding to the predicted maximum sunspot number of 118.1 $\pm$ 35. In Figure~4$b$ and 4$c$, 
we also show the estimated $\delta\nu$ for the current solar cycle using the predicted smoothed sunspot number
and 10.7 cm radio flux obtained from http://wwwssl.msfc.nasa.gov/ssl/pad/solar/predict.htm.

In summary, we have obtained  empirical relations between the
{\it change} in solar activity indices and the shift in p-mode
frequencies for cycle 22. From these empirical relations, we
have estimated p-mode frequency shift for the current cycle 23. It
will be of interest to see how our predicted value agrees with
the actual observed value during solar maximum in 2000. It is
further shown that these relations are
independent of the solar cycle and hence can be used to estimate
the change in frequency for past, present and future epochs,  if
the solar activity index is known.

\acknowledgements 

We thank T. Ata\c{c}, and  R.K. Ulrich for supplying us the
Flare index and MPSI data respectively.  The BBSO p-mode data
were acquired by Ken Libbrecht and Martin Woodard, Big Bear
Solar Observatory, Caltech. LOWL data were obtained from
(http://www.hao.ucar.edu/public/research/mlso/LowL/lowl.html),
NSO/Kitt Peak magnetic, and Helium measurements used here are
produced cooperatively by NSF/NOAO; NASA/GSFC and NOAA/SEL.
This work utilizes data obtained by the Global Oscillation
Network Group project, managed by the National Solar
Observatory, a Division of the National Optical Astronomy
Observatories, which is operated by AURA, Inc. under cooperative
agreement with the National Science Foundation. The data were
acquired by instruments operated by Big Bear Solar Observatory,
High Altitude Observatory, Learmonth Solar Obsrvatory, Udaipur
Solar Observatory, Instituto de Astrophsico de Canaris, and
Cerro Tololo Interamerican Observatory.  This work is partially
supported under the CSIR Emeritus Scientist Scheme and Indo-US
collaborative programme--NSF Grant INT-9710279.  

\end{article}

\begin{thebibliography}{}
\bibitem[\protect\citeauthoryear{Ata\c{c}}{1999}]{at99} Ata\c{c}, T.: 1999, private communication.
\bibitem[\protect\citeauthoryear{Bachmann and Brown}{1993}]{bb93}Bachmann, K. T. and Brown, T. M.: 1993, \apj {\bf 411}, L45.
\bibitem[\protect\citeauthoryear{Bhatnagar, Jain, and Tripathy}{1999}]{ab99}Bhatnagar, A., Jain, K., and
Tripathy, S. C.: 1999, \apj {\bf 521}, 885. 
\bibitem[\protect\citeauthoryear{Elsworth {\it et al}.}{1990}]{els90} Elsworth, Y., Howe, R., Isaak, G. R., Mcleod, C. P.,
~and New. R.: 1990, Nature {\bf 345}, 322.
\bibitem[\protect\citeauthoryear{Harvey}{1984}]{har84}Harvey, J. W.: 1984, in B. La Bonte, G. Chapman, H. Hudson,~and R. C. Wilson (eds.), 
Workshop on Solar Irradiance Variations on Active Region Time Scales, NASA CP-2310; NASA, Washington, p. 197.
\bibitem[\protect\citeauthoryear{Howe, Komm, and Hill}{1999}]{howe99}Howe, R., Komm, R., and Hill, F.: 1999, \apj {\bf 524}, 1084. 
\bibitem[\protect\citeauthoryear{Jain {\it et al}.}{1999}]{jain99}Jain, K., Tripathy, S. C., Kumar, B., and Bhatnagar, A. 1999, Bull. Astron.
Soc. India (in Press). 
\bibitem[\protect\citeauthoryear{Jim\'{e}nez-Reyes {\it et al}.}{1998}]{jim98}Jim\'{e}nez-Reyes, S. J., R\'{e}gulo, C , Pall\'{e}, P. L.,
and Roca Cort\'{e}s, T.: 1998, \aap {\bf 329}, 1119. 
\bibitem[\protect\citeauthoryear{Libbrecht and Woodard}{1990}]{lw90}Libbrecht, K. G.,
and Woodard, M. F.: 1990, in  Osaki, Y. and Shibahashi, H. (eds.), Progress of 
Seismology of the Sun and Stars, Springer, p. 145.
\bibitem[\protect\citeauthoryear{Pall\'{e}, R\'{e}gulo,~and Roca Cort\'{e}s}{1989}]{palle89}Pall\'{e}. P. L., R\'{e}gulo, C.,~and 
Roca Cort\'{e}s, T.: 1989, \aap  {\bf 224}, 253.
\bibitem[\protect\citeauthoryear{Rhodes {\it et al}.}{1993}]{rhodes93} Rhodes, E. J., Jr., Cacciani, A., 
Korzennik, S. G., and Ulrich, R. K.: 1993, \apj {\bf 406}, 714.
\bibitem[\protect\citeauthoryear{Ulrich }{1991}]{ulr91}Ulrich, R. K.: 1991, Adv. Space Res. {\bf 11(4)}, 217.
\bibitem[\protect\citeauthoryear{Woodard et al.}{1991}]{wod91}Woodard, M. F., Kuhn, J. R., Murray, N.,~and Libbrecht, K. G.: 1991, 
\apj {\bf 373}, L81.
\bibitem[\protect\citeauthoryear{Woodard and Noyes}{1985}]{wn85}Woodard, M. F.~and Noyes, R. W.: 1985, Nature {\bf 318}, 449.

\end{thebibliography}
\end{document}